\documentclass[preprintnumbers, floatfix, showkeys, preprintnumbers, letterpaper, twocolumn, superscriptaddress,nofootinbib]{revtex4-2}
\pdfoutput=1 
\usepackage{graphicx}
\usepackage{microtype}
\usepackage{amsmath}
\usepackage{amssymb}
\usepackage{subfigure}
\usepackage{url}
\usepackage{hyperref}
\usepackage{mathtools}
\usepackage{orcidlink}

\usepackage{xcolor}
\usepackage{color}
\usepackage{mathrsfs}
\usepackage{calrsfs}
\usepackage{amsfonts}
\usepackage{tabularx}
\usepackage{latexsym}
\usepackage{ragged2e}
\usepackage{epsfig}
\usepackage{textcomp}
\usepackage{float}

\usepackage{caption}
\DeclareCaptionJustification{justified}{\leftskip=0pt \rightskip=0pt \parfillskip=0pt plus 1fil}
\definecolor{vividviolet}{rgb}{0.62, 0.0, 1.0}
\definecolor{amaranth}{rgb}{0.9, 0.17, 0.31}
\definecolor{palatinateblue}{rgb}{0.15, 0.23, 0.89}
\definecolor{brightpink}{rgb}{1.0, 0.0, 0.5}
\definecolor{cornflowerblue}{rgb}{0.39, 0.58, 0.93}
\definecolor{deepcarminepink}{rgb}{0.94, 0.19, 0.22}
\definecolor{radicalred}{rgb}{1.0, 0.21, 0.37}

\hypersetup{ linktoc=all,
	colorlinks, linkcolor={palatinateblue},
	citecolor={brightpink}, urlcolor={amaranth}
}

\graphicspath{{Images/}}

\renewcommand{\d}[1]{\ensuremath{\operatorname{d}\!{#1}}}

\def\sideremark#1{\ifvmode\leavevmode\fi\vadjust{\vbox to0pt{\vss
			\hbox to 0pt{\hskip\hsize\hskip1em
				\vbox{\hsize1.3cm\tiny\raggedright\pretolerance10000
					\noindent #1\hfill}\hss}\vbox to8pt{\vfil}\vss}}}%
\def\beq{\begin{equation}}
\def\eeq{\end{equation}}

\begin{document}
\title{Kiselev Black Holes in Anti-de Sitter Spacetime Are Not Viable}

\author{Jiayi \surname{Xia}\orcidlink{0009-0003-4585-6208}}\email{xiajiayi@xbdx.wecom.work}
\affiliation{School of Physics, Northwest University, Xi'an 710127, China}

\author{Yen Chin \surname{Ong}\orcidlink{0000-0002-3944-1693}}\email{ongyenchin@nuaa.edu.cn}
\affiliation{Center for the Cross-disciplinary Research of Space Science and Quantum-technologies (CROSS-Q), College of Physics, Nanjing University of Aeronautics and Astronautics, \\29 Jiangjun Road, Nanjing City, Jiangsu Province 211106, China}\affiliation{Center for Gravitation and Cosmology, College of Physical Science and Technology, Yangzhou University, \\180 Siwangting Road, Yangzhou City, Jiangsu Province 225002, China}

\begin{abstract}
While the Kiselev black hole solution does not describe a perfect fluid quintessence, it is possible that dark energy fluid is not isotropic, and so one may hope that the solution may still describe at least locally a black hole embedded in such a dark energy fluid environment. Taking into account that the accelerated expansion of the Universe is caused by a dynamical dark energy fluid, and the actual underlying cosmological constant is negative as has been sometimes suggested, we show that the Kiselev solution is untenable as it is unstable against brane nucleation instability.
\end{abstract}

\maketitle

\section{Introduction: The Kiselev Black Hole and Dark Energy}\label{I}

The simplest Kiselev black hole solution \cite{0210040} is described by the spherically symmetric metric tensor\footnote{In this work we set $c=G=\hbar=1$.}
\begin{equation}
ds^2 = -f(r)dt^2 + f(r)^{-1} dr^2 + r^2(d\theta^2+\sin^2\theta d\varphi^2),
\end{equation}
with 
\begin{equation}
f(r)=1-\frac{2M}{r}-\frac{K}{r^{1+3w}},~~K>0.
\end{equation}
The case $K=0$ is just the usual Schwarzschild solution with mass $M$. The correction term has two parameters: the coefficient $K$ and the equation of state parameter $w$, whose interpretation will become obvious later.
This metric describes a black hole immersed in an environment with a fluid whose density $\rho$, radial pressure $p_r$, and tangential pressure $p_t$ satisfy \cite{0210040,1908.11058}:
\begin{equation}
\rho = -p_r = -\frac{3Kw}{8\pi r^{3(1+w)}},
\end{equation}
and 
\begin{equation}
p_t = -\frac{3Kw(1+3w)}{8\pi r^{3(1+w)}}.
\end{equation}
If we insist on positive energy density for the fluid $\rho > 0$, then with $K>0$, we must have $w<0$. Usually quintessence in cosmology satisfies $-1< w < -1/3$, so this is also the range typically considered in the study of Kiselev solution in the literature. The parameter $w$ is best seen as the equation of state of the averaged pressure fluid: $\bar{p}=w\rho$, where \cite{1908.11058}
\begin{equation}
\bar{p} = \frac{p_r+2p_t}{3} = -\frac{3Kw^2}{8\pi r^{3(1+w)}}.
\end{equation}

It must however be emphasized that the fluid in Kiselev spacetime is \emph{not} a perfect fluid because it is not isotropic \cite{1908.11058}. Despite the name ``quintessence'' used to describe such a fluid in the literature, it does not conform with the usual usage of the term in cosmology, which refers to an \emph{isotropic} scalar field with a timelike gradient. Nevertheless, there are models of dark energy in which the fluid is not isotropic\footnote{The conditions for such a fluid to serve as a dark energy are \cite{1009.4403}: (i) $\rho+p_r \geqslant 0$, (ii) $\rho+p_t \geqslant 0$, and (iii) $\rho + p_r + 2p_t < 0$. It can be checked that these are indeed satisfied for $w<-1/3$. See also \cite{0803.2508}.}. In such a case, at the cosmological scale, the expansion in different direction would accelerate at different rates, leading to an anisotropic universe (described by, e.g., Bianchi-type solutions \cite{0801.3676,1103.2658,2408.08740}). However, just like we can still use the Schwarzschild solution to study black holes whose angular momentum is negligible in our Universe locally, ignoring its expansion at cosmological scale, we can in principle also consider the Kiselev solution locally, even if at the largest scale the Universe is anisotropic. The question is whether such a solution is viable even for such a purpose.

In this work, we will argue that it is \emph{not} provided that the underlying true cosmological constant $\Lambda$ is negative. The possibility that $\Lambda < 0$ \cite{0403104,2212.00050,2211.12611,2307.12763,2506.04306,2410.05875} is not only theoretically viable, but also has been hinted by observations. Even if the cosmological constant is negative, there can be a temporary accelerated expansion phase due to a dark energy fluid. However, the presence of a negative $\Lambda$ introduces a new term in the metric, so that we have a Kiselev-anti-de Sitter metric, whose metric function reads
\begin{equation}\label{metricfunction}
f(r)=\frac{r^2}{L^2}+1-\frac{2M}{r}-\frac{K}{r^{1+3w}},~~K>0,
\end{equation}
where $L$ is the curvature length scale, related to the cosmological constant by $\Lambda=-3/L^2$. (It should be mentioned that some aspects of the Kiselev-AdS black holes have been studied in the literature \cite{2107.01282,2340-5}.) In anti-de Sitter spacetime, we may then take into account whether the solution is viable under brane nucleation instability. We provide a quick review of this method in Sec.(\ref{sec2}) below, and continue with our analysis in Sec.(\ref{sec3}), and finally conclude in Sec.(\ref{sec4}). 

\section{Brane Nucleation Instability}\label{sec2}

Start by Wick-rotating the black hole solution and impose a periodic boundary condition on the imaginary time direction as usual.
We first introduce the crucial quantity, the (Euclidean) brane action $\mathfrak{S}^E$, before explaining its physical significance. This action, in $D$-dimensional spacetime, is given by
\begin{flalign}
\mathfrak{S}^\text{E}\coloneq &~r^{D-2} \int \d \Omega_{D-2} \int \d \tau \sqrt{g_{\tau\tau}}\\ \notag &- \frac{D-1}{L} \int \d\tau \int \d \Omega_{D-2} \int^r_{r_\text{eh}} \d r' r'^{D-2} \sqrt{g_{\tau\tau}g_{r'r'}},
\end{flalign}
in which the metric component $g_{\tau\tau}$ is the Wick-rotated metric coefficient of $g_{tt}$. 
Here $r_\text{eh}$ denotes the (Euclidean) event horizon. The action is, up to a brane tension constant coefficient, that of a BPS (Bogomol'nyi–Prasad–Sommerfeld) probe brane that wraps around
the black hole at a constant coordinate radius $r$.
The normalized time coordinate $t/L$ is periodically identified with a periodicity $2\pi P$. The integral then gives 
\begin{equation}\label{actionF}
\mathfrak{S}^\text{E}= 2\pi P L \Omega_{D-2} \left(\underbrace{r^{D-2}\sqrt{g_{\tau\tau}}-\frac{r^{D-1}-r_\text{eh}^{D-1}}{L}}_{=:\mathcal{F}(r)}\right).
\end{equation}
Here $\Omega_{D-2}$ denotes the unit area of the $(D-2)$-dimensional sphere. For our case, it is just the unit area of $S^2$, i.e., $\Omega_2=4\pi$.
The action essentially measures the difference between the surface area of a probe brane and the volume (normalized by $1/L$) enclosed by it. If the normalized volume is larger than the area, the action becomes negative, which is problematic because the positivity of $\mathfrak{S}^\text{E}$ is a geometric consistency condition analogous to the ``isoperimetric inequality'' for asymptotically hyperbolic manifolds, as explained from various perspectives in the literature \cite{1310.6788,1311.4520,1411.1887,1504.07344,1602.07177,1610.07313}. See \cite{2208.13360} for more explanations and an application of this method in the context of Einstein-Gauss-Bonnet black holes. For applications in holography of quark-gluon plasma, see \cite{0905.1180,0910.4456}.

The Lorentzian picture is perhaps easier to appreciate: the pair production rate of brane-anti-brane pairs goes like $\exp(-\mathfrak{S}^\text{L})$, where $\mathfrak{S}^\text{L}$ is the Lorentzian brane action\footnote{The Lorentzian action may not be positive even if the Euclidean one is when the metric involves charges or angular momentum. In such cases we should require that both actions be positive \cite{1504.07344}.}. When the action becomes negative, the spacetime will nucleate a copious amount of branes, similar to how the Schwinger process produces charged particles when the critical field is reached. This brane nucleation becomes exponentially enhanced instead of suppressed indicates that the solution is not a stable one, non-perturbatively speaking \cite{0409242,1005.4439}. This is sometimes referred to as the ``Seiberg-Witten instability'' \cite{9903224}, and the unstable black holes are described as being ``fragile'' \cite{1008.0231}. 

\section{Analysis of the Brane Action}\label{sec3}
Since the factor $2\pi PL \Omega_{D-2}$ in Eq.(\ref{actionF}) is strictly positive, the sign of the action $\mathfrak{S}^\text{E}$ is the same as the sign of the
function $\mathcal{F}(r)$. Thus it suffices for us to study the behavior of this function.
In our case, for the 4-dimensional Kiselev-AdS black hole, we observe that
\begin{flalign}
\mathcal{F}(r) &= r^2\left(1+\frac{r^2}{L^2}-\frac{2M}{r}-\frac{K}{r^{1+3w}}\right)^{\frac{1}{2}} - \frac{r^3}{L} + \frac{r_\text{eh}^3}{L}. \notag \\ 
&= \frac{r^3}{L}\left(\frac{L^2}{r^2}+1-\frac{2ML^2}{r^3}-\frac{KL^2}{r^{3+3w}}\right)^{\frac{1}{2}} - \frac{r^3}{L} + \frac{r_\text{eh}^3}{L}. \notag \\ 
&=\frac{r^3}{L}\left(1-\frac{ML^2}{r^3}+\frac{L^2}{2r^2}-\frac{KL^2}{2r^{3+3w}}+\cdots\right) - \frac{r^3}{L} + \frac{r_\text{eh}^3}{L}
\notag \\
&=L\left(-M+\frac{r}{2}-\frac{K}{2r^{3w}}+\frac{r_\text{eh}^3}{L^2}\right)+\cdots.
\end{flalign}
In the third line, we assume that $r$ is sufficiently large to take the series expansion.
If the equation of state parameter is $w=-1+\varepsilon/3$ (with $\varepsilon \ll 1$ so that it is close to a constant positive cosmological constant), then 
\begin{equation}
\mathcal{F}(r) = L\left(-M+\frac{r}{2}-\frac{Kr^{3-\varepsilon}}{2}+\frac{r_\text{eh}^3}{L^2}\right)+\cdots.
\end{equation} 
There is a danger that the third term in the bracket can dominate for large $r$ for some values of $\varepsilon$, therefore it is indeed possible that the spacetime suffers from brane nucleation instability (recall that $K>0$). In fact, we see that in order for this to \emph{not} occur, the third term cannot be larger in magnitude compared to the second term $r/2$. This is equivalent to the requirement that $3-\varepsilon \leqslant 1$, i.e., $\varepsilon \geqslant 2$. Instead of $\varepsilon \ll 1$, we need it to be sufficiently large. This in turn implies the inequality
\begin{equation}\label{cond}
w=-1+\frac{\varepsilon}{3} \geqslant -\frac{1}{3}.
\end{equation}
\emph{This stability condition is exactly the opposite of the requirement for a dark energy fluid} ($w<-1/3$)!

This calculation establishes our claim that the Kiselev-AdS black holes are untenable as a stable solution (as a putative low energy solution of string theory). Of course, in asymptotically locally anti-de Sitter spacetimes, black holes with different topology are permitted, and the horizon can be flat\footnote{Holographic superconductor for such a black hole was investigated in \cite{1206.2069}.} or negatively curved. The metric function is generalized by replacing the 1 in Eq.(\ref{metricfunction}) with $k$, with $k=+1,0,-1$ corresponding to positively curved, flat, and negatively curved horizons, respectively. It is clear that $k=0$ and $k=-1$ would render the brane action more negative\footnote{The $\Omega_{D-2}$ in the action is replaced by $\Omega_{k,D-2}$. For example, for a cubic torus with periodicity $2\pi K$ on each copy of its $S^1$, we have $\Omega_{0,2}=4\pi^2K^2$.}. In any case, if one is motivated by cosmology of the actual Universe, then we should only consider the $k=1$ case since that is the sort of black holes that were observed by astronomers.

For a consistency check of the stability condition Eq.(\ref{cond}), we note that the case with $\omega=1/3$ corresponds to the \emph{Euclidean} AdS-Reissner-Nordstr\"om black hole if $K=Q^2$ (recall that in order to ensure the Euclidean potential 1-form to be everywhere regular, Wick rotation not only calls for the complexification of time by $t\mapsto it$, but also that of the electric charge: $Q \mapsto -iQ$ \cite{1504.07344}, thus $K=Q^2>0$ is correct. For the Lorentzian case, we need $K=-Q^2<0$ instead \cite{1908.11058}). In \cite{1504.07344} the planar ($k=0$) case is discussed; the associated brane action is always non-negative (things become more delicate if the magnetic charge is present). The $k=1$ case is also stable.

\begin{figure}[!h]
\centering
\includegraphics[width=0.48\textwidth]{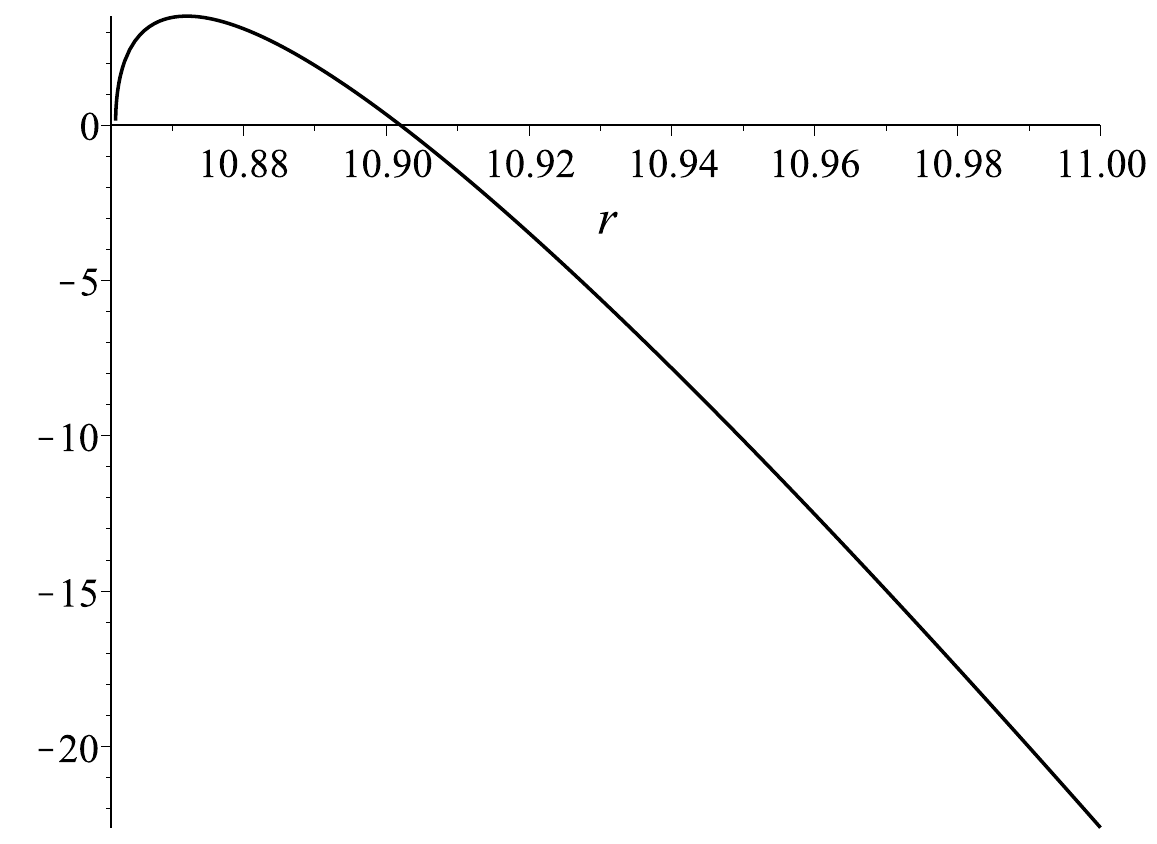}
\caption{The normalized brane action, $\mathcal{F}(r)=\mathfrak{S}^\text{E}/8\pi^2 P L$, for Kiselev-AdS black hole with $K=L=1,M=10,\varepsilon=0.003$. The action is not bounded from below. \label{fig1}}
\end{figure}

Though it is not necessary, 
we can perhaps better appreciate the behavior of the brane action by numerically examine some explicit examples without taking the series approximation. In Fig.(\ref{fig1}), we show the function $\mathcal{F}(r)$ for parameter values $K=1,L=1,M=10,\varepsilon/3=0.001$. The horizon can be solved as $r_\text{eh}\approx 10.862$. The brane action first increases from its zero value at the horizon, then turns around and eventually becomes zero again at $r\approx 10.902$ before becoming negative. 

\begin{figure}[!h]
\centering
\includegraphics[width=0.48\textwidth]{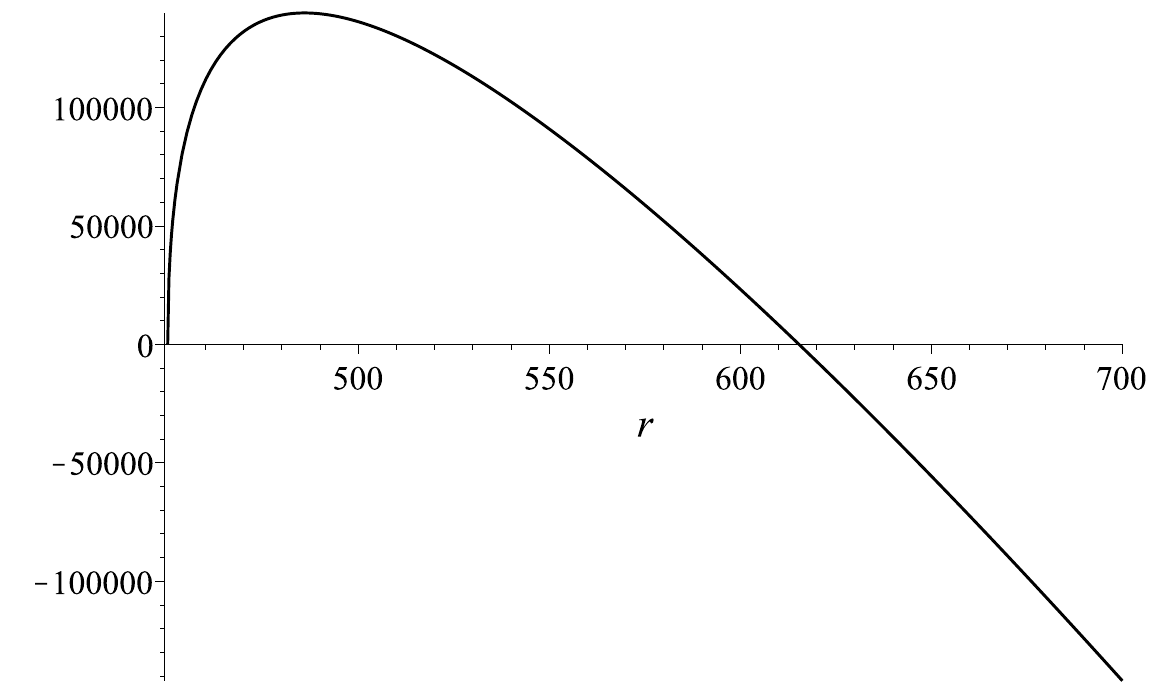}
\caption{The normalized brane action for Kiselev-AdS black hole, $\mathcal{F}(r)=\mathfrak{S}^\text{E}/8\pi^2 P L$, with $K=1,L=100,M=10,\varepsilon=1.5$. The action is not bounded from below. \label{fig2}}
\end{figure}

For larger values of $L$, say $L=100$, the black hole horizon is located at extremely large value of $r$ if $\varepsilon$ is small. To obtain a better plot we can consider, for example, $\varepsilon/3=0.5$, i.e. $\varepsilon=1.5$. Then the horizon is located at $r_\text{eh} \approx 450.115$, and the brane action becomes negative starting at around $r_\text{eh}\approx 615.38$. This is shown in Fig.(\ref{fig2}).

It should be mentioned that another bound comes from the requirement that the spacetime should be asymptotically AdS. This means that the metric function $f(r)$, Eq.(\ref{metricfunction}), should grow at most like $r^2$ asymptotically. The Kiselev term grows as $r^{-1-3\omega}$. Therefore one must impose the inequality $2> -1-3\omega$, which gives the lower bound
\begin{equation}
-1< \omega. 
\end{equation}
This, however, does not give us anything new since it is consistent with the quintessence condition $-1 < \omega < -1/3$.

\section{Conclusion: Kiselev-Anti-de Sitter Black Holes Are Not Viable}\label{sec4}

In this article, we have shown that Kiselev-AdS black holes are not stable against brane nucleation. Such a black hole, while being a perfectly well-defined solution in classical general relativity, is not a consistent low-energy solution of quantum gravity (at least, in string theory). This is analogous to the fact that not all effective field theories can be embedded into string theory, as discussed in depth in the contexts of ``Swampland Conjecture'' \cite{0605264,1903.06239}. Of course, there is a difference: the brane action analysis reveals whether a classical \emph{solution} in AdS is consistent, instead of whether a theory is consistent. 

It is interesting to emphasize that this instability sets in precisely when $w\leqslant -1/3$, the exact range required for the fluid to account for the accelerated expansion of the Universe. However, we should be careful about what this actually implies. It does \emph{not} indicate that there can be no black hole solution with the anisotropic fluid with $\omega < -1/3$ in asymptotically AdS spacetime, only that the solution is not given by the one examined here. Indeed, the copious brane production is expected to back-react on the geometry and it would likely settle down into a new configuration (admittedly, whether this is a new black hole or a horizonless object is unknown). 

To conclude, if indeed our Universe has a negative cosmological constant, and the current accelerated expansion is provided by an anisotropic dark energy fluid, the local geometry in the vicinity of a black hole cannot be described by the Kiselev-AdS solution.


\begin{thebibliography}{99}

\bibitem{0210040}
Valeri V. Kiselev, ``Quintessence and Black Holes'', {\hypersetup{urlcolor=vividviolet}\href{https://iopscience.iop.org/article/10.1088/0264-9381/20/6/310}{Class. Quant. Grav. \textbf{20} (2003) 1187}}, \href{https://arxiv.org/abs/gr-qc/0210040v3}{[arXiv:gr-qc/0210040]}.

\bibitem{1908.11058}
Matt Visser, ``The Kiselev Black Hole Is neither Perfect Fluid, nor Is It Quintessence'', {\hypersetup{urlcolor=vividviolet}\href{https://iopscience.iop.org/article/10.1088/1361-6382/ab60b8}{Class. Quant. Grav. \textbf{37} (2020) 4, 045001}}, \href{https://arxiv.org/abs/1908.11058v1}{[arXiv:1908.11058 [gr-qc]]}.

\bibitem{1009.4403}
R. Chan, M.F.A. da Silva, P. Rocha, ``Gravastars and Black Holes of Anisotropic Dark Energy'', {\hypersetup{urlcolor=vividviolet}\href{https://link.springer.com/article/10.1007/s10714-011-1178-6}{Gen. Rel. Grav. \textbf{43} (2011) 2223}}, \href{https://arxiv.org/abs/1009.4403v2}{[arXiv:1009.4403 [gr-qc]]}.

\bibitem{0803.2508}
R. Chan, M. F. A. da Silva, J. F. Villas da Rocha, ``On Anisotropic Dark Energy'', {\hypersetup{urlcolor=vividviolet}\href{https://www.worldscientific.com/doi/abs/10.1142/S0217732309028692}{Mod. Phys. Lett. A \textbf{24} (2009) 1137}}, \href{https://arxiv.org/abs/0803.2508}{[arXiv:0803.2508 [gr-qc]]}.

\bibitem{0801.3676}
Tomi Koivisto, David F. Mota, ``Anisotropic Dark Energy: Dynamics of Background and Perturbations'', {\hypersetup{urlcolor=vividviolet}\href{https://iopscience.iop.org/article/10.1088/1475-7516/2008/06/018}{JCAP \textbf{06} (2008) 018}}, \href{https://arxiv.org/abs/0801.3676}{[arXiv:0801.3676 [astro-ph]]}.

\bibitem{1103.2658}
L. Campanelli, P. Cea, G. L. Fogli, L. Tedesco, ``Anisotropic Dark Energy and Ellipsoidal Universe'', {\hypersetup{urlcolor=vividviolet}\href{https://www.worldscientific.com/doi/abs/10.1142/S021827181101927X}{Int. Jour. Mod. Phys. D \textbf{20} (2011) 1153}}, \href{https://arxiv.org/abs/1103.2658}{[arXiv:1103.2658 [astro-ph.CO]]}.

\bibitem{2408.08740}
Anshul Verma, Pavan K. Aluri, David F. Mota, ``Anisotropic Universe With Anisotropic Dark Energy'', {\hypersetup{urlcolor=vividviolet}\href{https://journals.aps.org/prd/abstract/10.1103/PhysRevD.111.083508}{Phys. Rev. D \textbf{111} (2025) 8, 083508}}, \href{https://arxiv.org/abs/2408.08740}{[arXiv:2408.08740 [astro-ph.CO]]}.


\bibitem{0403104}
Brett McInnes, ``Quintessential Maldacena-Maoz Cosmologies'', {\hypersetup{urlcolor=vividviolet}\href{https://doi.org/10.1088/1126-6708/2004/04/036}{JHEP \textbf{04} (2004) 036}}, \href{https://arxiv.org/abs/hep-th/0403104}{[arXiv:hep-th/0403104]}.

\bibitem{2212.00050}
Stefano Antonini, Petar Simidzija, Brian Swingle, Mark Van Raamsdonk, Chris Waddell, ``Accelerating Cosmology From $\Lambda < 0$ Gravitational Effective Field Theory'',  {\hypersetup{urlcolor=vividviolet}\href{https://link.springer.com/article/10.1007/JHEP05(2023)203}{JHEP \textbf{05} (2023) 203}}, \href{https://arxiv.org/abs/2212.00050}{[arXiv:2212.00050 [hep-th]]}.

\bibitem{2211.12611}
Mark Van Raamsdonk, ``Cosmology Without Time-Dependent Scalars Is Like Quantum Field Theory Without RG Flow'', \href{https://arxiv.org/abs/2211.12611}{[arXiv:2211.12611 [hep-th]]}.

\bibitem{2307.12763}
Shahnawaz A. Adil, Upala Mukhopadhyay, Anjan A. Sen, Sunny Vagnozzi, ``Dark Energy in Light of the Early Jwst Observations: Case for a Negative Cosmological Constant?'', {\hypersetup{urlcolor=vividviolet}\href{https://iopscience.iop.org/article/10.1088/1475-7516/2023/10/072}{JCAP \textbf{10} (2023) 072}}, \href{https://arxiv.org/abs/2307.12763}{[arXiv:2307.12763 [astro-ph.CO]]}.


\bibitem{2410.05875}
Prasanta Sahoo, Nandan Roy, Himadri Shekhar Mondal, ``Quintessence Scalar Field and Cosmological Constant: Dynamics of a Multi-Component Dark Energy Model'', {\hypersetup{urlcolor=vividviolet}\href{https://link.springer.com/article/10.1007/s10714-025-03372-7}{Gen. Rel. Grav. \textbf{57} (2025) 2, 38}}, \href{https://arxiv.org/abs/2410.05875}{[arXiv:2410.05875 [gr-qc]]}.

\bibitem{2506.04306}
Hao Wang, Yun-Song Piao, ``Can the Universe Experience an Ads Landscape Since Matter-Radiation Equality?'', \href{https://arxiv.org/abs/2506.04306}{[arXiv:2506.04306 [gr-qc]]}.

\bibitem{2107.01282}
Valeria Ramírez, L. A. López, Omar Pedraza, V. E. Ceron, ``Scattering and Absorption Sections of Schwarzschild-Anti de Sitter With Quintessence'', {\hypersetup{urlcolor=vividviolet}\href{https://cdnsciencepub.com/doi/10.1139/cjp-2021-0269}{Can. J. Phys. \textbf{100} (2022) 2, 112}}, \href{https://arxiv.org/abs/2107.01282v2}{[arXiv:2107.01282 [gr-qc]]}.


\bibitem{2340-5}
B. Malakolkalami, K. Ghaderi, ``Schwarzschild-Anti de Sitter Black Hole With Quintessence'', {\hypersetup{urlcolor=vividviolet}\href{https://link.springer.com/article/10.1007/s10509-015-2340-5}{Astrophys. Space. Sci. \textbf{357} (2015) 112}}.

\bibitem{1310.6788}
Frank Ferrari, ``Gauge Theories, D-Branes and Holography'', {\hypersetup{urlcolor=vividviolet}\href{https://doi.org/10.1016/j.nuclphysb.2014.01.007}{Nucl. Phys. B \textbf{80} (2014) 247}}, \href{https://arxiv.org/abs/1310.6788}{[arXiv:1310.6788 [hep-th]]}.

\bibitem{1311.4520}
Frank Ferrari, ``D-Brane Probes in the Matrix Model'', {\hypersetup{urlcolor=vividviolet}\href{https://linkinghub.elsevier.com/retrieve/pii/S0550321313006068}{Nucl. Phys. B \textbf{880} (2014) 290}}, \href{https://arxiv.org/abs/1311.4520}{[arXiv:1311.4520 [hep-th]]}.

\bibitem{1411.1887}
Frank Ferrari, Antonin Rovai, ``Holography, Probe Branes and Isoperimetric Inequalities'', {\hypersetup{urlcolor=vividviolet}\href{https://www.sciencedirect.com/science/article/pii/S0370269315004268?via\%3Dihub}{Phys. Lett. B \textbf{747} (2015) 212}}, \href{https://arxiv.org/abs/1411.1887}{[arXiv:1411.1887 [hep-th]]}.

\bibitem{1504.07344}
Brett McInnes, Yen Chin Ong, ``When Is Holography Consistent?'', {\hypersetup{urlcolor=vividviolet}\href{https://linkinghub.elsevier.com/retrieve/pii/S0550321315002412}{Nucl. Phys. B \textbf{898} (2015) 197}}, \href{https://arxiv.org/abs/1504.07344}{[arXiv:1504.07344 [hep-th]]}.


\bibitem{1602.07177}
Frank Ferrari, Antonin Rovai, ``Gravity and On-Shell Probe Actions'', {\hypersetup{urlcolor=vividviolet}\href{https://link.springer.com/article/10.1007\%2FJHEP08\%282016\%29047}{JHEP \textbf{08} (2016) 047}}, \href{https://arxiv.org/abs/1602.07177}{[arXiv:1602.07177 [hep-th]]}.

\bibitem{1610.07313}
Brett McInnes, ``Isoperimetric Inequalities and Magnetic Fields at CERN'', Asia-Pacific Mathematics Newsletter \textbf{6} (2016) 10, \href{https://arxiv.org/abs/1610.07313}{[arXiv:1610.07313 [hep-th]]}.

\bibitem{2208.13360}
Yen Chin Ong, ``Holographic Consistency and the Sign of the Gauss-Bonnet Parameter'', {\hypersetup{urlcolor=vividviolet}\href{https://www.sciencedirect.com/science/article/pii/S0550321322002905}{Nucl. Phys. B \textbf{984} (2022) 115939}}, \href{https://arxiv.org/abs/2208.13360}{[arXiv:2208.13360 [hep-th]]}.

\bibitem{0905.1180}
Brett McInnes, ``Bounding the Temperatures of Black Holes Dual to Strongly Coupled Field Theories on Flat Spacetime'', {\hypersetup{urlcolor=vividviolet}\href{https://iopscience.iop.org/article/10.1088/1126-6708/2009/09/048}{JHEP \textbf{09} (2009) 048}}, \href{https://arxiv.org/abs/0905.1180}{[arXiv:0905.1180 [hep-th]]}.

\bibitem{0910.4456}
Brett McInnes, ``Holography of the Quark Matter Triple Point'', {\hypersetup{urlcolor=vividviolet}\href{https://www.sciencedirect.com/science/article/abs/pii/S0550321310000994?via\%3Dihub}{Nucl. Phys. B \textbf{832} (2010) 323}}, \href{https://arxiv.org/abs/0910.4456}{[arXiv:0910.4456 [hep-th]]}.


\bibitem{0409242}
Matthew Kleban, Massimo Porrati, Raul Rabadan, ``Stability in Asymptotically AdS Spaces'', {\hypersetup{urlcolor=vividviolet}\href{https://iopscience.iop.org/article/10.1088/1126-6708/2005/08/016}{JHEP \textbf{08} (2005) 016}}, \href{https://arxiv.org/abs/hep-th/0409242}{[arXiv:hep-th/0409242]}.

\bibitem{1005.4439}
Jos\'e L. F. Barb\'on, Javier Mart\'inez-Mag\'an, ``Spontaneous Fragmentation of Topological Black Holes'', {\hypersetup{urlcolor=vividviolet}\href{https://link.springer.com/article/10.1007\%2FJHEP08\%282010\%29031}{JHEP \textbf{08} (2010) 031}}, \href{https://arxiv.org/abs/1005.4439}{[arXiv:1005.4439 [hep-th]]}.

\bibitem{9903224}
Nathan Seiberg, Edward Witten, ``The D1/D5 System And Singular CFT'', {\hypersetup{urlcolor=vividviolet}\href{https://iopscience.iop.org/article/10.1088/1126-6708/1999/04/017}{JHEP \textbf{04} (1999) 017}}, \href{https://arxiv.org/abs/hep-th/9903224}{[arXiv:hep-th/9903224]}.

\bibitem{1008.0231}
Brett McInnes, ``Fragile Black Holes'', {\hypersetup{urlcolor=vividviolet}\href{https://www.sciencedirect.com/science/article/abs/pii/S0550321310004372?via\%3Dihub}{Nucl. Phys. B \textbf{842} (2011) 86}}, \href{https://arxiv.org/abs/1008.0231}{[arXiv:1008.0231 [hep-th]]}.


\bibitem{1206.2069}
Songbai Chen, Qiyuan Pan, Jiliang Jing, ``Holographic Superconductors in Quintessence AdS Black Hole'', {\hypersetup{urlcolor=vividviolet}\href{https://iopscience.iop.org/article/10.1088/0264-9381/30/14/145001}{Class. Quant. Grav. \textbf{30} (2013) 145001}}, \href{https://arxiv.org/abs/1206.2069}{[arXiv:1206.2069 [gr-qc]]}.


\bibitem{0605264}
Hirosi Ooguri, Cumrun Vafa, ``On the Geometry of the String Landscape and the Swampland'', {\hypersetup{urlcolor=vividviolet}\href{https://linkinghub.elsevier.com/retrieve/pii/S0550321306008455}{Nucl. Phys. B \textbf{766} (2007) 21}}, \href{https://arxiv.org/abs/hep-th/0605264}{[arXiv:hep-th/0605264]}.

\bibitem{1903.06239}
Eran Palti, ``The Swampland: Introduction and Review'', {\hypersetup{urlcolor=vividviolet}\href{https://onlinelibrary.wiley.com/doi/abs/10.1002/prop.201900037}{Fortsch. Phys. \textbf{67} (2019) 6, 1900037}}, \href{https://arxiv.org/abs/1903.06239}{[arXiv:1903.06239 [hep-th]]}.



\end{thebibliography}
\end{document}